\documentclass[prl,aps,amsfonts,amssymb,twocolumn,floats,showpacs]{revtex4}
\usepackage{graphicx}

\begin{document}
\title{Critical level-statistics for weakly disordered graphene}

\author{H. Amanatidis, I. Kleftogiannis, D.E. Katsanos and S.N. Evangelou\footnote{e-mail:sevagel@cc.uoi.gr}}
\affiliation{Department of Physics, University of Ioannina,
Ioannina 45110, Greece}

\begin{abstract}
In two dimensions chaotic level-statistics is expected for massless Dirac fermions in the presence of disorder. For weakly disordered graphene flakes with zigzag edges the obtained level-spacing distribution in the Dirac region is neither chaotic (Wigner) nor localized (Poisson) but similar to that at the critical point of the Anderson metal-insulator transition. The quantum transport in finite graphene can occur via critical edge states as in topological insulators, for strong disorder the Dirac region vanishes and graphene behaves as ordinary Anderson insulator.

\end{abstract}


\maketitle
\section{}

\par
\medskip
Graphene, a monolayer of graphite with carbons arranged in a two-dimensional honeycomb lattice has attracted great interest since its pioneering fabrication\cite{r1,r2,r3,r4}. Its high mobility, current-carrying capacity and thermal conduction makes it a good candidate to replace silicon in future nanoelectronics\cite{r5}. Graphene is a zero gap semiconductor, its Fermi surface consists of two non-equivalent Dirac points with linear energy dispersion in the Dirac region. The sublattice (chiral) symmetry of graphene guarantees the presence of the Dirac region where electrons behave as massless Dirac fermions, limits backscattering, introduces Andreev reflection and Klein tunnelling, etc\cite{r6}. In order to control its gapless nature graphene was folded with cyclic boundary conditions in the short axis to form long cylinders (nanotubes\cite{r7}), or long narrow strips with open perpendicular edges (nanoribbons\cite{r8,r9}). In nanoribbons if the lattice geometry contains zigzag (zz) edges\cite{r8,r9,r10,r11} the sublattice symmetry, which is responsible for the Dirac equation, guarantees the presence of $E=0$ edge states tied to the boundary\cite{r8,r9}. We have studied Anderson localization\cite{r12} in finite disordered graphene samples of various shapes known as graphene flakes\cite{r13}, that is in confined lattices fabricated from graphene. 

\par
\medskip
In two-dimensional(2D) disordered systems, with time-reversal and spin-rotation, all states are localized even for very small disorder, while in 3D localization occurs only above the Anderson metal-insulator transition\cite{r14}. In 2D, however, the states have localization lengths which often exceed the system size. In other words, for weak disorder the states behave as diffusive, they extend from one end of the sample to the other, and the energy levels give chaotic (Wigner) level-spacing distribution\cite{r15}. The observed level-repulsion and spectral rigidity are the main characteristics of quantum chaos, the Poisson exponential law which denotes localization requires unrealistically large system sizes to occur. In weakly disordered graphene flakes with zz edges for open boundary conditions (bc) the energy levels for a wide disorder region ($W<W_{c}$ in Fig.1) show unusual quantum chaos, they resemble a gas of solitons\cite{r16}, etc. The corresponding level-spacing distribution is neither Wigner (chaotic) nor Poisson (localized), is critical as at the 3D Anderson transition\cite{r17}. The energy spectrum, obtained by numerical diagonalization under the constraint of open bc, gives our main result: weakly disordered graphene flakes with zz edges in the Dirac region show a novel kind of quantum chaos ($W<W_{c}$ in Fig.1) with critical level-statistics, similar to what is known at the Anderson transition\cite{r14}. 

\begin{figure}[hbtp]
\centering
\includegraphics[scale=0.5]{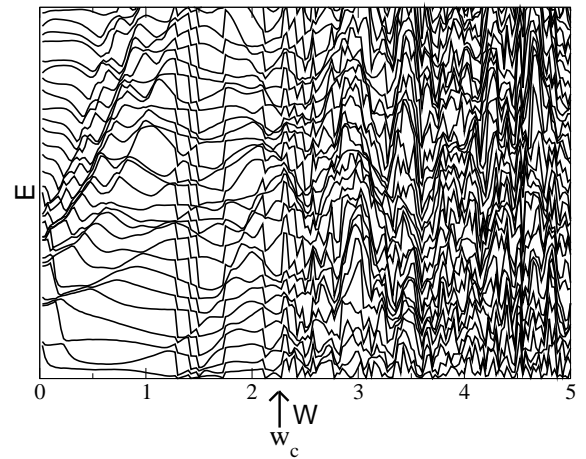}
\caption{The energy levels $E$ {\it vs.} the strength of disorder $W$ for a quarter circle graphene flake of $9691$ sites. For $W<W_{c}$ the level-statistics is critical and for $W>W_{c}$ is localized (Poisson). The levels were unfolded by removing variations of the density of states $\rho(E)$ averaged over $5000$ realizations of disorder. In this case $W_{c}\simeq 2.3$.}
\end{figure}

\par
\medskip
Our calculations are done for the nearest neighbour tight-binding Hamiltonian
\begin{equation}
      H=\sum_{i}\varepsilon_{i}c_{i}^{\dag}c_{i} -
      \sum_{<i,j>}\gamma _{i,j}(c_{i}^{\dag}c_{j}+c_{j}^{\dag}c_{i}),
\end{equation}
$c_{i} (c_{i}^{\dag})$ annihilates(creates) an electron at A or B
site $i$ of the honeycomb lattice, the diagonal disorder $\varepsilon_{i}$ is constant in the range $[-W/2,+W/2]$, $W$ is its  strength and ${<i,j>}$ denotes nearest neighbours with hopping matrix elements $\gamma_{i,j}=1$. For $W=0$ the band structure displays two non-equivalent valleys (act as pseudospin states) related by time-reversal\cite{r5,r6} and in the Dirac region (at long length scales) Eq.(1) can be replaced by the continuous Dirac equation\cite{r6}. The short-range diagonal disorder via $\epsilon_{i}$ causes intervalley scattering and mixes the two-valleys which leads to localization. The decoupling of valleys by breaking time-reversal symmetry occurs for a smooth long-range disordered potential, this involves scattering within a single valley and gives weak antilocalization familiar from spin-orbit coupling\cite{r18}. The disorder $\varepsilon_{i}$ mixes valleys and destroys A,B sublattice symmetry. 

\par
\medskip
We have cut the honeycomb lattice in various shapes, circular, stadium, square, etc. The shape of the flake turns out to be irrelevant for our study but the perimeter, as we shall see, it is not. In our computations the brick-wall lattice is mostly used, however we find no significant differences from the honeycomb lattice itself. The eigenvalues of Eq.(1) in the Dirac region are obtained via Lanczos numerical diagonalization, building a statistical ensemble of random $H$ matrices for every $W$. For $W=0$ quantum chaos for stadium flakes and integrability for circular flakes is expected, while if the lattice geometry contains zz edges the sublattice symmetry guarantees the presence of edge states.  The edge states due to the symmetry of the bulk belong to one type (A or B) sublattice and they are protected against disorder by the lattice topology which is reflected in the Hilbert space structure. In the honeycomb lattice the edge states arise from quantum interference of an incoming wave from one bond which splits into two. In the presence of weak disorder the edge states spread almost uniformly in the Dirac region\cite{r10}. For zero disorder from the tight-binding equations of the bulk\cite{r8,r9} their amplitude $\Psi (m)$ is nonzero only for $m=0$ and they are localized at the boundary having a fractal dimension of $1$. For nonzero $W$ they penetrate into the bulk and their fractal dimension varies. These massless degrees of freedom at the edges for finite $W$ are remnants of the sublattice symmetry which is broken by the diagonal disorder of Eq.(1). 

\begin{figure}[hbtp]
\centering
\includegraphics[scale=0.5]{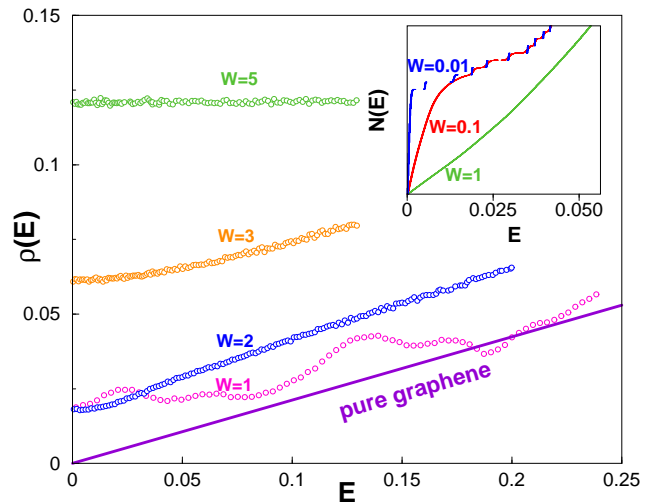}
\caption{The averaged density of states $\rho(E)$ for circular graphene flakes with disorder $W$ obtained from more than one million eigenvalues of Eq.(1). For $W=1,2$ the edge states lie close to $E=0$ where $\rho(E)$ is nonzero and almost constant. For strong disorder ($W=5$) their proportion becomes vanishingly small while in the opposite limit of pure ($W=0$) graphene: $\rho(E)=\frac{1}{\pi \sqrt{3}} |E|$ (continuous line) is linear. Inset: the averaged integrated density of states for weaker disorder $W=0.01, 0.1$ the  $N(E)=\int_{0^{+}}^E\!\rho(E')\,\,\mathrm{d}E'$ (the $E=0$ states are not included) shows minigaps due to low $\rho(E)$ values, while for  higher $W$ the structure disappears.}
\end{figure}

\par
\medskip
In Fig.2 the averaged density of states $\rho(E)$ for various $W$ is shown. In the absence of disorder ($W=0$) the zz edges  contribute to zero energy, if they disappear the edge states also disappear. Their total number remains fixed for nonzero disorder\cite{r10}, e.g. for $W=1,2$ the majority is in the almost constant $\rho(E)$ region of Fig.2. Moreover, for weaker disorder $W=0.01, 0.1$  minibands and minigaps develop in the spectrum around the discrete $W=0$ states, this is due (inset of Fig.2) to the small\ $\rho(E)$ and the finite lattice. We have also computed $N(E)$ which counts the number of states from $E=0^{+}$ to $E$. In a log-log plot we find constant $N(E)/E$ vs $E$ which implies constant $\rho(E)$ for very low $E$, for slightly higher $E$ the $\rho(E)$ decreases for $W<W_{c}$ (due to a minigap) and it increases for $W>W_{c}$ (no minigaps). For strong disorder ($W=5$) states fill up the Dirac region and make $\rho(E)$ constant. 

\par
\medskip
We have examined in detail the perimeter of the considered flakes, identified each type of lattice edge and its contribution to $\rho(E)$. We find armchair and zz edges (the dangling bonds are rare), the most frequent edges are the zz\cite{r10,r11}. The ratio of zz to armchair edges for circular flakes is $\sim 3.8$ and it varies linearly with the averaged flake radius $R$ (the linear curve showed more oscillations for the zz edges). The ratio of zz edges over the total number of sites tends to zero inversely proportional to $R$, since their number is $\sim R$ and the total number of sites is proportional to the area of the flake $\sim R^{2}$, the  density of states $\rho(0)$ reaches a maximum before it vanishes as $1/R$ for large $R$. The disorder shifts the edge states away from zero energy without changing their total number\cite{r10}.

\begin{figure}[hbtp]
\centering
\includegraphics[scale=0.4]{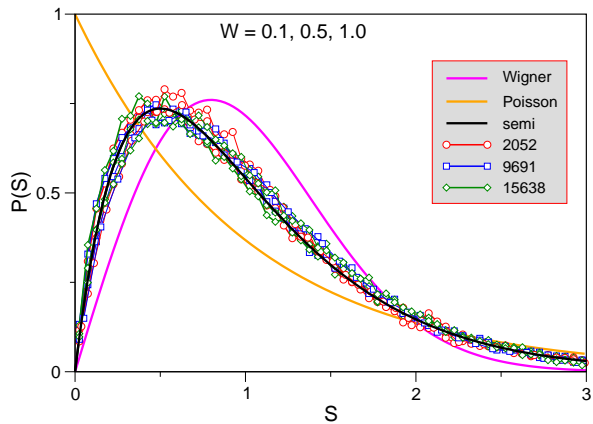}
\caption{The calculated spacing distribution $P(S)$ of the first two positive energies ($S=E_{2}-E_{1}$) for $W<W_{c}$ for circular quarter graphene flakes in the new phase $W<W_{c}$, sizes $N=2052$, $9691$, $15638$ and $50000$ realisations of disorder. The data are independent of size and are described by the intermediate  semi-Poisson  $P(S)=4S\exp(-2S)$ distribution (black line).}
\end{figure}

\par
\medskip
Let us now discuss the eigenvalues of Eq.(1).
We examined the level-statistics of the unfolded energy spectra via the nearest spacing distribution $P(S)$ which can distinguish between chaotic (Wigner) and localized (Poisson)\cite{r14}. In Fig.3 the obtained $P(S)$ of the first two positive levels ($S=E_{2}-E_{1}$) is shown for the critical region of $W<W_{c}$. The $P(S)$ is independent of size and fits into a curve which has mixed chaotic and localized  character, it is chaotic for small spacing $S$ and localized for large $S$\cite{r17}, it is described by $P(S)=4S\exp(-2S)$ and belongs to the semi-Poisson family\cite{r19}. The critical $P(S)$ interpolates between chaotic and localized limits: $P(S)\sim S^{\beta}$ for $S\to 0$ and $P(S)\sim \exp(-(1+\beta)S)$ for $S\to \infty$, the universality class index $\beta=1$, for broken  time-reversal symmetry via a magnetic field $\beta =2$\cite{r14} and for broken spin rotation via spin-orbit coupling $\beta=4$\cite{r18}.

\par
\medskip
For pure graphene ($W=0$) we find Poisson level-statistics in the Dirac region since the integrated density of states $\it{N}(E\sim \sqrt{k_{x}^{2}+k_{y}^{2}})\sim E^{2}$ gives integrable unfolded levels $\alpha n^{2}+m^{2}$, $n,m=1, 2,...$, $\alpha\sim 1$ for a square sample. For graphene with zz edges the critical distribution replaces Poisson for $W<W_{c}$(Fig.1) which can describe only larger $S$,  while for small $S$ the $P(S)$ is Wigner-like.  The obtained critical distribution interpolates between chaotic and the localized limits (apart from 3D critical systems it was found for pseudo-integrable billiards, etc.\cite{r19}). In graphene localization is easier than other 2D due to its small coordination, this counterbalances the chiral direction of electrons in the Dirac region which favours absence of localization.  The effect of edges vanishes by taking periodic boundary conditions, as it was seen in previous numerical studies of level-statistics\cite{r21,r22}. For strong disorder the Dirac region vanishes altogether and Poisson is obtained for localized states as for zero disorder. The flow of $P(S)$ towards Poisson as the size increases also occurs for the honeycomb  lattice but for a smaller $W$ than for the square. The topological effects found in Fig.1 vanish for infinite size and/or strong disorder unlike in topological insulators\cite{r23}.

\par
\medskip
\begin{figure}[hbtp]
\centering
\includegraphics[scale=0.4]{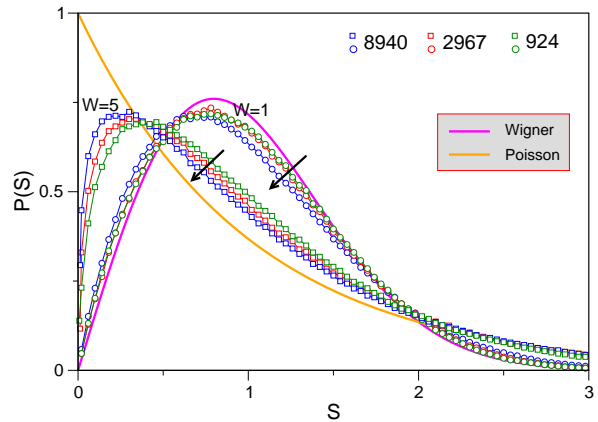}
\caption{The flow of the level-spacing distribution $P(S)$ towards Poisson as the system size increases. The data for stronger disorder $W>W_{c}$, $W=1, 5$ are obtained from stadium quarter flakes of sizes $N=924$, $2967$, and $8940$, for $50000$ realisations of disorder and energies in the window $[0,0.12]$ of the Dirac region. The two arrows indicate the flow towards the localized (Poisson) limit as the system size increases, the approach although slow is much faster than that for the square lattice. The Poisson limit verifies that in infinite graphene all states are localized.}
\end{figure}

\par
\medskip
\begin{figure}[hbtp]
\centering
\includegraphics[scale=0.4]{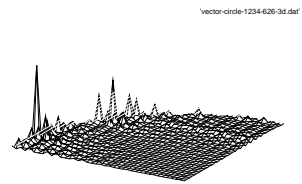}
\caption{A critical edge state at nonzero $E=0.14$ for a quarter circular flake with disorder $W=2$ ($W<W_{c}$). Its fractal dimension is close to $1$.}
\end{figure}

\par
\medskip
Our work was partially motivated by the fabrication of graphene quantum dots for diameters ranging from 40 to 100 nanometers\cite{r13}. For small flakes the experimental level-statistics is increasingly non-Poisson while the very small flakes of a few nm width remained always conductive. The obtained critical level-statistics is different from that of conventional 2D disordered systems. We find that the sample size is also crucial, the fast or slow approach to localization depends on the kind of lattice, in graphene for strong disorder $W>W_{c}$ the approach to localization is faster than in other 2D (Fig.4). For weak disorder $W<W_{c}$ the level-statistics depends on the lattice termination, e.g for a nanoribbon of arbitrary orientation which has a majority of zz edges\cite{r11} critical level-statistics should also appear. The  edge states are prominent only for small size and weak disorder and are intimately connected to the small $\rho(E)$ in the Dirac region. It is also interesting to ask whether the critical level-statistics can give conductive transport. 

\par
\medskip
The edge states in flakes and nanoribbons are protected from disorder, they cannot be deleted without changing the topology of the Hilbert space. Edge states of topological origin are found for bilayer graphene with subgap conductance \cite{r24}, are  seen in microwave realizations of nanoribbons\cite{r25}, they are spatially resolved  with scanning tunnelling microscopy\cite{r26}, etc. The new phase found for $W<W_{c}$ also shows the absence of weak localization\cite{r1,r2,r3,r4} in graphene. The boundary geometry plays important role, the critical $P(S)$ is independent of size but it depends on bc as at the 3D Anderson transition\cite{r27}. In the presence of a magnetic field transport  via protected edge states occurs in the integral quantum Hall effect\cite{r14}. It would be interesting to examine our findings in the presence of a magnetic field, for other types of disorder, such as edge disorder, resonant impurities, vacancies, adding the true spin of electrons and electron-electron interactions.

\par
\medskip
The level-statistics of weakly disordered graphene flakes as a function of short-ranged disorder is not chaotic as in other 2D where the localization length is usually much larger than the system size. The realistic edge structure for $W<W_{c}$ makes the spectrum in the Dirac region intermediate between chaotic and integrable, similar to that at the critical point of the Anderson metal-insulator transition. Graphene is different than other weakly disordered 2D, its topological behaviour is also suggested by the finite conductivity $\sigma \sim e^{2}/h$ at the Dirac point\cite{r6}. On the basis of our results a mapping of electrons to the classically integrable Calogero-Moser model of interacting particles is possible\cite{r16}, a kink mechanism relates chaotic behavior to classical equations. In graphene the edge is important and when combined with the low $\rho(E)$ gives the novel critical regime of Fig.1. The smaller coordination number of the lattice affects a graphene flake, we find smaller and smaller $W_{c}$ by increasing the flake's size and in the infinite size limit the spectrum becomes gapless. In other words, graphene has no intrinsic topological order but a pseudogapped phase for weak disorder and finite size. 
 
\par
\medskip
In conclusion, graphene for weak disorder can conduct via its edges. A sharp transition at $W_{c}$ is found (Fig.1) which distinguishes between critical ($W<W_{c}$) and localized ($W>W_{c}$) states. This is shown in the Dirac region where the density of states is low and minibands appear with edge states. Our results are a signature that weakly disordered graphene is topological insulator-like, its conduction depends on the topology of the perimeter.

\end{document}